\documentclass[aps,prl,twocolumn,superscriptaddress,showpacs]{revtex4}

\usepackage{times}
\usepackage{mathptmx}
\usepackage{amsmath}
\usepackage{amsfonts}
\usepackage{amssymb}
\usepackage{graphicx}
\usepackage{hyperref}

\newcommand{\ud}{\mathrm{d}}

\def\addUMD{Maryland Center for Fundamental Physics \& Joint Space-Science Institute, Department of Physics, University of Maryland, College Park, MD 20742, USA}
\def\addSouth{School of Mathematics, University of Southampton, Southampton SO17 1BJ, United Kingdom}
\def\addCITA{Canadian Institute for Theoretical Astrophysics, University of Toronto, Toronto, Ontario M5S 3H8, Canada}
\def\addYuk{Yukawa Institute for Theoretical Physics, Kyoto University, Kyoto 606-8502, Japan}

\begin{document}

\title{Periastron Advance in Black Hole Binaries}

\author{Alexandre Le Tiec}
\affiliation{\addUMD}

\author{Abdul H. Mrou{\'e}}
\affiliation{\addCITA}

\author{Leor Barack}
\affiliation{\addSouth}

\author{Alessandra~Buonanno}
\affiliation{\addUMD}

\author{Harald P. Pfeiffer}
\affiliation{\addCITA}

\author{Norichika Sago}
\affiliation{\addYuk}

\author{Andrea Taracchini}
\affiliation{\addUMD}

\begin{abstract}
The general relativistic (Mercury-type) periastron advance is calculated here for the first time with exquisite precision in full general relativity. We use accurate numerical relativity simulations of spinless black hole binaries with mass ratios $1/8 \leqslant m_1/m_2 \leqslant 1$ and compare with the predictions of several analytic approximation schemes. We find the effective-one-body model to be remarkably accurate, and, surprisingly, so also the predictions of self-force theory [replacing $m_1/m_2 \to m_1 m_2/(m_1+m_2)^2$]. Our results can inform a universal analytic model of the two-body dynamics, crucial for ongoing and future gravitational-wave searches.
\end{abstract}


\maketitle

{\it Introduction.} The anomalous rate of Mercury's
perihelion advance was originally recognized in 1859 by the astronomer
Urbain Le Verrier. For the first time, Newton's law of universal
gravitation could not be reconciled with observation. Treating Mercury
as a test body in free fall in the gravitational field generated by
the mass $M_\odot$ of the Sun, Einstein derived the lowest order (weak-field)
general relativistic angular advance per orbit \cite{Ei.15}
\begin{equation}\label{Einstein}
\Delta \Phi = \frac{6\pi G\, M_\odot}{c^2\,a\,(1 - e^2)} \, ,
\end{equation}
where $a$ and $e$ are the semi-major axis and eccentricity of
Mercury's orbit, respectively. Equation \eqref{Einstein} perfectly accounted for
the observed discrepancy of $\sim 43"$ per century, thus providing the first
successful test of general relativity. More recently, the same effect---but with a much larger
amplitude, of order a few degrees per year---has been observed in the orbital motion of
binary pulsars \cite{St.03}. Today, the exciting prospects of observing gravitational waves
from the inspiral and merger of compact binaries, using interferometric detectors 
like LIGO or Virgo, provide a modern context for the problem of relativistic 
periastron advance, and a motivation to go far beyond Einstein's 
weak-field test-particle approximation.

In this Letter we restrict our attention to binaries composed of two black holes. Their orbital dynamics can be analyzed using several approximation schemes in general relativity: post-Newtonian expansions \cite{Bl.06}, black hole perturbation theory \cite{Po.04}, and the effective-one-body model \cite{BuDa.99}. It can also be studied using fully nonlinear numerical relativity (NR). While NR can now routinely perform accurate binary black hole simulations \cite{Ce.al.10}, approximation methods remain valuable given the high computational cost of these simulations, and their restricted utility when the mass ratio is too extreme. It is important to assess the predictions of the various approximations against the NR benchmark, since (i) it allows crucial cross-validation tests, (ii) it helps delineate the respective domains of validity of each method, and (iii) it can inform the development of a universal semi-analytical model of the binary dynamics.
 
Neglecting radiation reaction, the motion of two non-spinning black holes on a generic eccentric orbit involves two frequencies: the radial frequency (or mean
motion) $\Omega_r$, and the averaged angular frequency $\Omega_\varphi$, defined by
\begin{equation}\label{Omegas}
\Omega_r = \frac{2 \pi}{P} \, , \quad 
\Omega_\varphi = \frac{1}{P} \int_0^P \! \dot{\varphi}(t) \, \ud t = K \, \Omega_r \, ,
\end{equation}
where $P$ is the radial period, i.e. the time interval between two successive periastron passages, $\dot{\varphi} = \ud \varphi / \ud t$ is the time derivative of the orbital phase $\varphi(t)$, and $\Delta \Phi / (2 \pi) = K-1$ is the fractional advance of the
periastron per radial period. In the circular orbit limit, the relation between $K =
\Omega_\varphi / \Omega_r$ and $\Omega_\varphi$ is coordinate
invariant (for a large class of physically reasonable coordinate
systems), and therefore provides a natural reference for comparing between the 
predictions of the analytical and numerical methods currently available.

In this Letter we present new accurate NR simulations starting
at lower orbital frequencies than in previous work
\cite{Lo.al.10,Sp.al.11,Bu.al.11}.
We outline the respective computations of the invariant relation
$K(\Omega_\varphi)$ in numerical relativity, post-Newtonian theory,
the effective-one-body formalism, and black hole perturbation theory.
We then perform an extensive comparison which, for the first time,
(i) encompasses all of these methods, and (ii) focuses on the
orbital dynamics of the binary, rather than the asymptotic
gravitational waveform. We also discuss the implications for the modelling of coalescing 
compact binaries. (We henceforth set $G = c = 1$.)

{\it Numerical Relativity.} The periastron advance of non-spinning
black hole binaries was estimated for the first time in
general relativistic numerical simulations in \cite{Mr.al.10}. In the
present work, we improve considerably on the accuracy of these
calculations. Our results are based on new and longer simulations of the late stage of the
inspiral of black hole binaries, using the Spectral Einstein Code {\tt SpEC}
\cite{SpECwebsite,Bo.al2.07}, with mass ratios $q \equiv
m_1 / m_2$ between $1{:}1$ and $1{:}8$, and eccentricities $e$ in the
range $[0.0015,0.023]$. These runs are summarized in Table~\ref{tab:K-fit},
and will be described in detail elsewhere \cite{Bu.al.11,Mr.al.11}.
(Ref.~\cite{Mr.al.10} discusses the definition of $e$ in these simulations.)

We compute $\Omega_\varphi$ and $\Omega_r$ using the
orbital frequency $\Omega(t)$ extracted from the motion of
the apparent-horizon centers (in harmonic coordinates): let $\mathbf {c}_{i}(t)$ be the
coordinates of the center of each black hole, and define their
relative separation $\mathbf{r}= \mathbf{c}_1-\mathbf{c}_2$; then
$\Omega = \vert \mathbf{r} \times \dot{\mathbf{r}} \vert / r^2$, where
the Euclidean cross product and norm are used. The frequency
$\Omega(t)$ can be written as the sum of a secular piece (given by the
average frequency $\Omega_\varphi$) and a small oscillatory
remainder---both of which drift slowly in time due to
radiation reaction. To compute $K_{\rm NR}$ at some coordinate time $T$, we choose a 
time interval of width $W\times 2\pi/\Omega(T)$, centered on $T$,
and fit $\Omega(t)$ to the model $\Omega(t)=
p_0 (p_1-t)^{p_2} + p_3 \cos \big[p_4 + p_5 {(t-T)}+ p_6
(t-T)^2\big]$, where the $p_i$'s are fitting parameters. We then write
$\Omega_\varphi(T) = p_0(p_1-T)^{p_2}$ and $\Omega_r(T)=p_5$, compute
the ratio $K_\text{NR}(T)=\Omega_\varphi(T)/\Omega_r(T)$, and hence
obtain $K_\text{NR}$ as a function of $\Omega_\varphi$. Finally,
we fit $ K_\text{NR}(\Omega_\varphi)$ to a smooth quadratic polynomial using
\begin{equation}\label{fitting}
K_\text{NR} = \left[ a_0+a_1 (m \Omega_{\varphi}) + a_2 (m \Omega_{\varphi})^2 \right] K_\text{Schw} \, ,
\end{equation}
where $m = m_1 + m_2$ is the total mass of the binary. The results of the fits are given in Table~\ref{tab:K-fit}. For convenience, the numerical periastron advance $K_\text{NR}$ is normalized by the test-particle result $K_\text{Schw}$, which is known in closed form as~\cite{DaSc.88,Cu.al.94} $K_\text{Schw} = (1-6x)^{-1/2}$, where $x = (m \Omega_\varphi)^{2/3}$ is the usual dimensionless coordinate invariant post-Newtonian parameter.

\begin{table}
\begin{tabular}{ccccccccc}
\hline\hline
$q$ & $d/m$ & $e$ & $N_\text{orb}$ & $a_0$ & $a_1$ & $a_2$ & $m\Omega_i$ & $m\Omega_f$\\
\hline
$1$   & 19 & 0.021  & 34 & 0.9949 &  0.589  & -79.1  & 0.0111 & 0.0312\\
$2/3$ & 18 & 0.023  & 27 & 0.9950 &  0.573  & -75.9   & 0.0129 & 0.0316\\
$1/3$ & 14 & 0.002  & 29 & 0.9821 &  1.692   & -87.1   & 0.0181 & 0.0313\\
$1/5$ & 14 & 0.008  & 23 & 0.9879 &  1.154  & -62.8   & 0.0183 & 0.0361\\
$1/6$ & 13 & 0.015  & 20 & 0.9890 &  1.071  & -57.0  & 0.0204 & 0.0333\\
$1/8$ & 13 & 0.0015 & 24 & 1.0028 & -0.099 & -26.8 & 0.0197 & 0.0355\\
\hline\hline
\end{tabular}
\caption{\label{tab:K-fit} Simulation parameters. Here $q \equiv m_1/m_2$, $m \equiv m_1+m_2$, $d$ is the initial coordinate separation, $e$ the initial eccentricity, and $N_\text{orb}$ the total number of orbits in the simulation. The fitting parameters $\{a_0,a_1,a_2\}$ [cf. Eq.~\eqref{fitting}] are computed for the restricted frequency range $\Omega_i \leqslant \Omega_\varphi \leqslant \Omega_f$.}
\end{table}

The variance in the numerical data for various window sizes $W$ provides an estimate of the error in $K_\text{NR}$. We point out that the finite (non-zero) eccentricity in the NR simulations introduces a small error, since we are interested in the $e\to 0$ limit. However, as the leading-order result~\eqref{Einstein} suggests, and calculations at higher post-Newtonian (PN) orders confirm, this error scales like $e^2$, which in our simulations is always $\lesssim 5 \times 10^{-4}$, and decreasing monotonically with time.

\begin{figure}[t!]
	\includegraphics[scale=0.48]{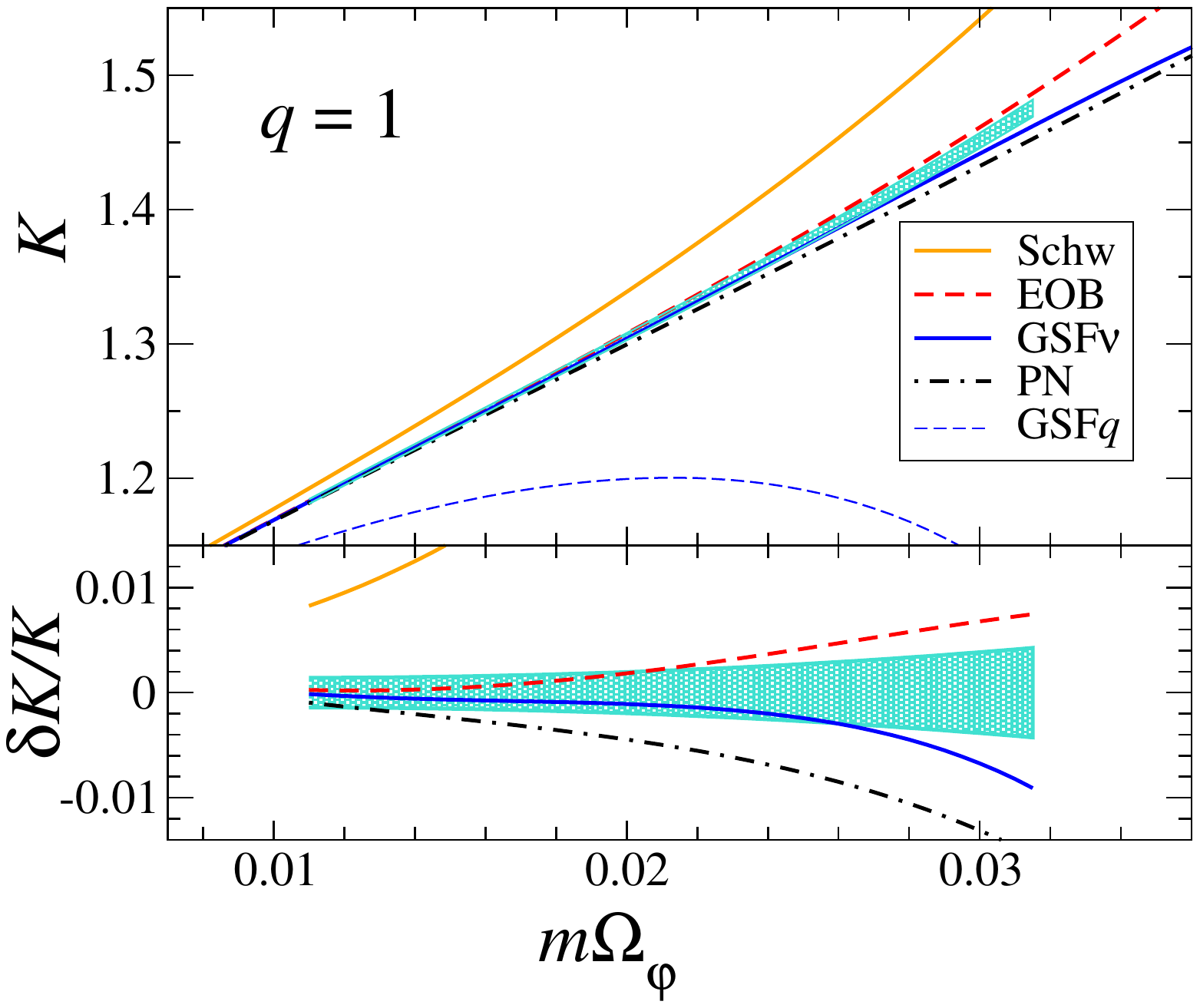}
	\caption{\footnotesize The periastron advance $K$ of an equal mass black hole binary, in the limit of zero eccentricity, as a function of the orbital frequency $\Omega_\varphi$ of the circular motion.  The NR results are indicated by the cyan-shaded region. The PN and EOB results are valid at 3PN order. The lower panel shows the relative difference $\delta K/K \equiv (K-K_{\rm NR})/K_{\rm NR}$.}
	\label{fig:q=1}
\end{figure}

\begin{figure}[t!]
	\includegraphics[scale=0.48]{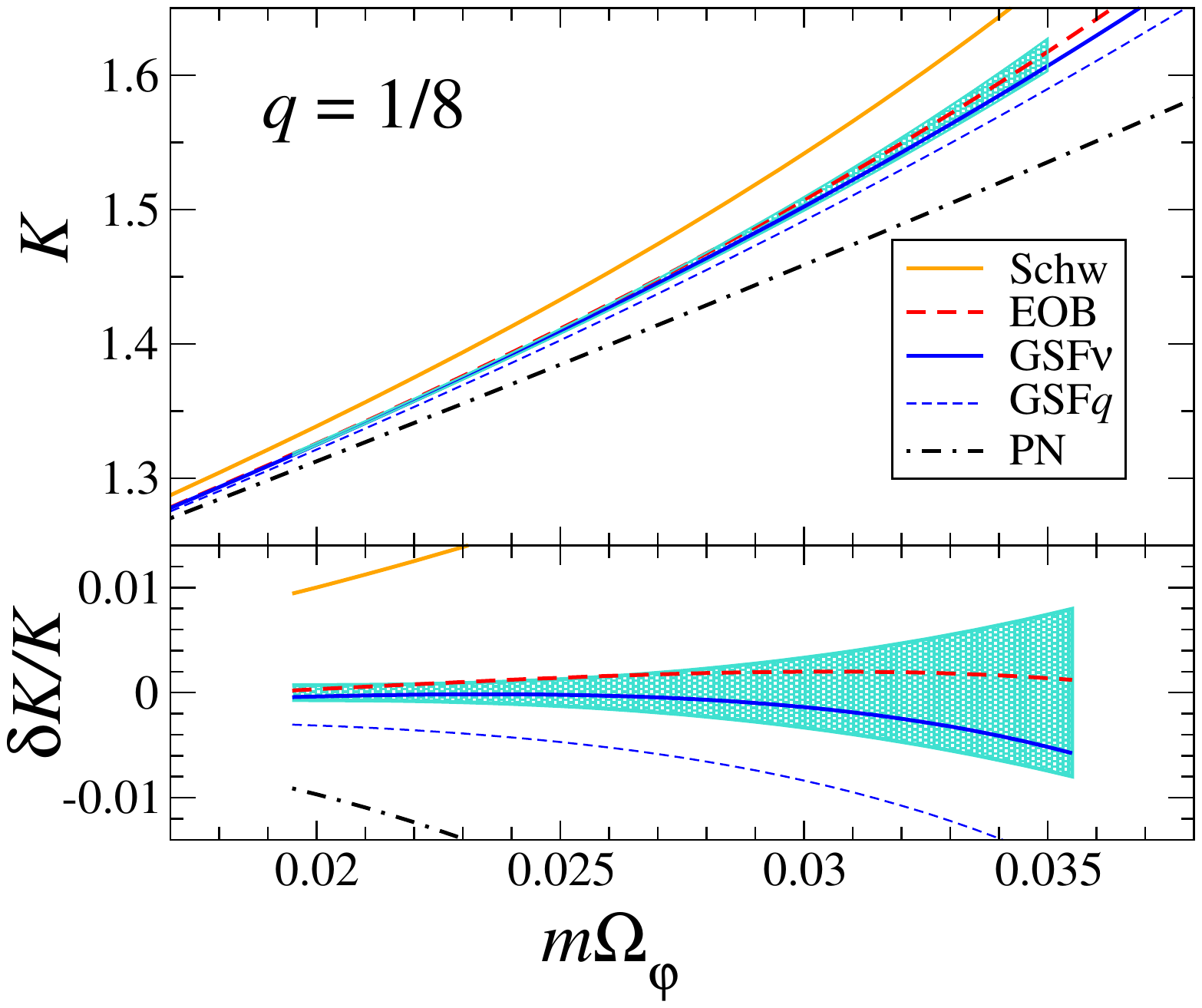}
	\caption{\footnotesize Same as in Fig.~\ref{fig:q=1}, but for a mass ratio $q={1}/{8}$. Note that for an orbital frequency $m \Omega_\varphi \sim 0.03$, corresponding to a separation $r \sim 10 m$, the periastron advance reaches half an orbit per radial period.}
	\label{fig:q=8}
\end{figure}

\begin{figure}[t!]
	\includegraphics[scale=0.475]{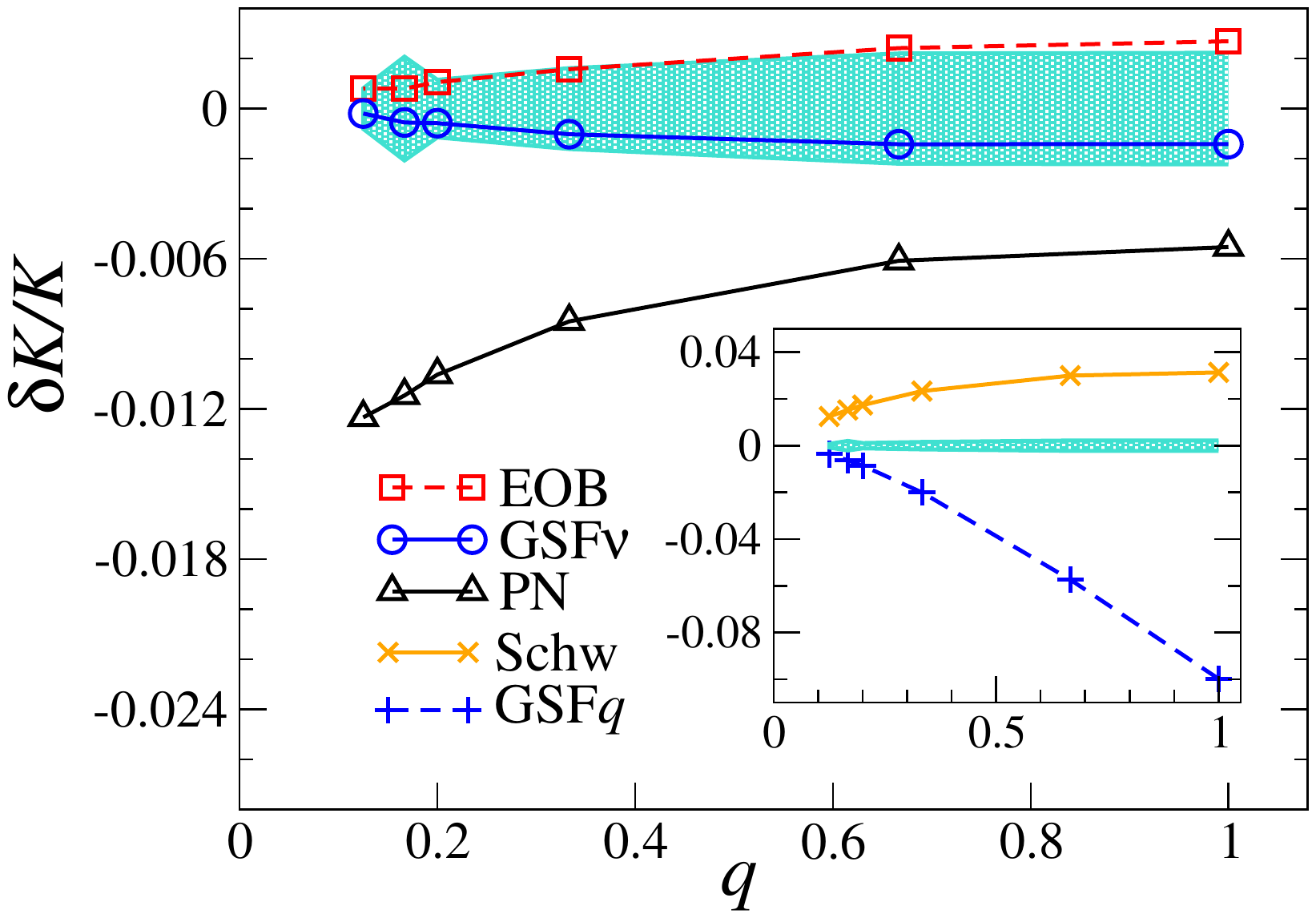}
	\caption{\footnotesize The relative difference $\delta K/K = (K - K_\text{NR}) / K_\text{NR}$ as a function of the mass ratio $q$, for $m \Omega_\varphi = 0.022$. The PN and EOB results are valid at 3PN order. The shaded area marks the error margin of the NR data. The results are qualitatively identical and quantitatively similar for other values of $\Omega_\varphi$.}
	\label{fig:Omega=0.022}
\end{figure}

The numerical data form the basis for our comparisons. We will now discuss the different approximation schemes in turn, summarizing the results in Figs.~\ref{fig:q=1} and \ref{fig:q=8} (showing $K$ as a function of frequency for two fixed mass ratios), and Fig.~\ref{fig:Omega=0.022} (showing $K$ as a function of mass ratio for a given frequency).

{\it Post-Newtonian Theory.} Einstein's result \eqref{Einstein} was
generalized to arbitrary masses $m_1$ and $m_2$ by
Robertson~\cite{Ro.38}. Following the discovery of binary pulsars in
the 1970s, an improved modelling of the orbital dynamics of these
compact binaries was required, leading to the extension of this 1PN
result to 2PN order~\cite{DaSc.88}. [As usual we
refer to $n$PN as the order equivalent to terms $\mathcal{O}(c^{-2n})$
in the equations of motion beyond the Newtonian acceleration.] More
recently, the need for extremely accurate gravitational-wave templates
modelling the inspiralling phase of coalescing compact binaries motivated
the computation of the equations of motion through 3PN order. 
These results allowed also the calculation of the periastron advance
at the 3PN accuracy for eccentric orbits \cite{Da.al.00}.

For quasi-circular orbits, combining Eqs.~(5.8) and (5.25) of
Ref.~\cite{Da.al.00}, we obtain the 3PN-accurate expression of $K$ as 
\begin{align}\label{3PN}
	K_\text{3PN} &=  1 + 3 x + \bigg( \frac{27}{2} - 7 \nu \biggr) \, x^2 \nonumber \\ &\hspace{-0.45cm} + \biggl( \frac{135}{2} - \biggl[ \frac{649}{4} - \frac{123}{32} \pi^2 \biggr] \nu + 7 \nu^2 \biggr) \, x^3 + \mathcal{O}(x^4) \, .
\end{align}
The symmetric mass ratio $\nu \equiv m_1 m_2 / m^2$ is such that $\nu =
{1}/{4}$ for an equal mass binary, and $\nu \rightarrow 0$ in the
extreme mass ratio limit. The term $\propto \nu^2$ in Eq.~\eqref{3PN}, which is a 3PN effect,
contributes less than $1\%$ to $K_\text{3PN}$, for all mass
ratios. This suggests that the exact $K$ may be well approximated by
a linear function of $\nu$. Figures~\ref{fig:q=1}--\ref{fig:Omega=0.022} show a good agreement between the 3PN and NR results for $q=1$, with $\lesssim 1\%$ relative difference even at the high-frequency end. However, the performance of the PN approximation deteriorates with decreasing $q$.

{\it Effective-One-Body (EOB).} The EOB formalism \cite{BuDa.99}
maps the conservative part of the PN dynamics of a compact
binary system onto the dynamics of a test particle of
reduced mass $\mu \equiv m \nu = m_1 m_2 / m$ in a time-independent and
spherically symmetric effective metric $\ud s^2_\text{eff} = -
A(r;\nu) \, \ud t^2 + B(r;\nu) \, \ud r^2 + r^2 (\ud \theta^2 +
\sin^2{\theta} \, \ud \varphi^2)$, 
which reduces to the Schwarzschild metric of a black hole of mass $m$ in the limit $\nu \rightarrow 0$. The expansions of the EOB potentials $A$ and $\bar{D} \equiv (A
B)^{-1}$ in terms of the Schwarzschild-like coordinate $u = m / r$ are known through 3PN order as \cite{BuDa.99,Da.al3.00} $A = 1 - 2 u + 2 \nu \, u^3 + \left( \frac{94}{3} - \frac{41}{32} \pi^2 \right) \nu \, u^4 + \mathcal{O}(u^5)$, and $\bar{D} = 1 + 6 \nu \, u^2 + (52 - 6 \nu) \, \nu \, u^3 + \mathcal{O}(u^4)$. To enforce the presence of an EOB innermost stable circular orbit (ISCO), Ref.~\cite{Da.al3.00} suggested replacing $A$ by its Pad{\'e} approximant of order $(1,3)$, $A_P = (1 + a u)/(1 + b u + c u^2 + d u^3)$, whose Taylor series coincides with the known 3PN result. 

From the recent analysis of slightly eccentric orbits in the EOB
formalism \cite{Da.10}, the effective-one-body prediction for the
periastron advance in the limit of zero eccentricity is given by
\begin{equation}
\label{EOB}
	K_\text{EOB} = \sqrt{\frac{A'_P(u)}{\bar{D}(u) \Delta(u)}} \, ,
\end{equation}
where $A'_P = \ud A_P / \ud u$, and $\Delta = A_P A'_P
+ 2 u (A'_P)^2 - u A_P A''_P$ vanishes at the EOB ISCO. To
obtain the invariant relation $K_\text{EOB}(x)$, one needs
to compute $u$ given $x$, which we do here numerically (for
any given $\nu$) from the expression of the EOB Hamiltonian restricted
to circular orbits, and Hamilton's equations of
motion~\cite{Da.10}. The resulting curves are displayed in red in
Figs.~\ref{fig:q=1}--\ref{fig:Omega=0.022}. For $q=1$ and $2/3$, the 
EOB(3PN) prediction \eqref{EOB} is within the numerical error up to
$m \Omega_\varphi \sim 0.022$. For all the other mass ratios, the EOB(3PN) 
result is within the numerical error at all frequencies.
When using the EOB potential $A(u)$ with 4PN and 5PN terms calibrated
to a set of highly accurate unequal mass non-spinning binary black
hole simulations \cite{Pa.al.11}, the EOB prediction is 
within the numerical error at all frequencies and for all mass ratios considered. This remarkable agreement 
could be attributed in part to the ``pole-like'' structure at the EOB ISCO in Eq.~\eqref{EOB},
which is absent from the standard PN result \eqref{3PN}.

{\it Perturbation Theory and the Gravitational Self-Force.} 
Extreme mass ratio inspirals (EMRIs) of compact objects into massive black holes, for which $m_2 \gg m_1$, are important sources of low-frequency gravitational radiation for future space-based detectors. Modelling the dynamics of these systems requires going beyond the geodesic approximation, by taking into account the back-reaction effect due to the interaction of the small object with its own gravitational perturbation.
This ``gravitational self-force'' (GSF) effect has recently been computed for generic (bound)
geodesic orbits  around a Schwarzschild black hole \cite{BaSa.09,BaSa.10,BaSa.11}. In particular, the
$\mathcal{O}(q)$ correction to the test-mass result $K_\text{Schw}$ has been derived \cite{Ba.al.10}.
This calculation determined (numerically) the term $\rho(x)$ in the function $W\equiv 1/K^2 = 1-6x + q \, \rho(x) + \mathcal{O}(q^2)$. The results are well 
fitted (at the $10^{-5}$ level) by the rational function $\rho = 14 x^2 (1 + \alpha x)/(1 + \beta x + \gamma x^2)$, with $\alpha =
12.9906$, $\beta = 4.57724$, and $\gamma = -10.3124$. (This model improves upon the model of Ref.~\cite{Ba.al.10}; it is based on a much 
denser sample of GSF data points in the relevant frequency range.) In terms of the quantity $K$ we have 
\begin{equation}
\label{GSFq}
K^q_\text{GSF} = \frac{1}{\sqrt{1-6x}} \left[ 1 - \frac{q}{2} \frac{\rho(x)}{1-6x} + \mathcal{O}(q^2) \right] .
\end{equation}
We used this expression, with the above analytic fit for $\rho(x)$, to produce the dashed blue curves in Figs.~\ref{fig:q=1}--\ref{fig:Omega=0.022}.

Since $\rho(x) > 0$ for all stable circular orbits, the $\mathcal{O}(q)$ GSF decreases the rate of precession. Note that the formal divergence of $K^q_\text{GSF}$ at the ISCO limit ($x\to 1/6$) is simply a consequence of the fact that $\Omega_{r}$ vanishes there (by definition), while $\Omega_{\varphi}$ remains finite. This divergence might explain why the convergence of the standard PN series seems to deteriorate with decreasing $q$ \cite{Bl.02}, as also illustrated by our results (cf.\ Fig.~\ref{fig:Omega=0.022}). We remind the reader that Eq.~\eqref{GSFq} captures only the {\em conservative} effect of the GSF, and has a limited physical relevance near the ISCO, where the actual dynamics transitions from an adiabatic quasi-circular inspiral (driven by the dissipative piece of the GSF) to a direct plunge \cite{BuDa.00,OrTh.00}.

We now turn to discuss one of the most striking findings of our study. Since $q$ and $\nu = q / (1+q)^2$ coincide at leading order, namely $q = \nu + \mathcal{O}(\nu^2)$, we may recast Eq.\ \eqref{GSFq} as
\begin{equation}
\label{GSFnu}
K^\nu_\text{GSF} = \frac{1}{\sqrt{1-6x}} \left[ 1 - \frac{\nu}{2} \frac{\rho(x)}{1-6x} + \mathcal{O}(\nu^2) \right] ,
\end{equation}
which, unlike $K^q_\text{GSF}$, is symmetric under $m_1\leftrightarrow m_2$. The solid blue curves in Figs.~\ref{fig:q=1}--\ref{fig:Omega=0.022} show $K^\nu_\text{GSF}$. Remarkably, while the agreement between $K^q_\text{GSF}$ and $K_\text{NR}$ becomes manifest only at sufficiently small $q$ (as expected), $K^\nu_\text{GSF}$ appears to agree extremely well with $K_\text{NR}$ at {\em all} mass ratios. This suggests that the substitution $q\to\nu$ amounts to an efficient ``resummation'' of the $q$-expansion, to the effect that much of the functional form $K(x)$ is captured by the $\mathcal{O}(\nu)$ term, even for large $q$.

A few heuristic explanations for this behavior may be suggested. (i) As mentioned earlier, quadratic corrections in $\nu$ enter the PN expression for $K$ only at 3PN [recall Eq.~\eqref{3PN}], and account for less than $1\%$ of $K$ at this order. This implies that the linear-in-$\nu$ approximation must be very accurate, at least at small frequencies. (ii) The true function $K(x;m_1,m_2)$ must be invariant under exchange $m_1\leftrightarrow m_2$. The expansion in $\nu$, $K_{\rm GSF}^\nu$, satisfies this symmetry by definition of $\nu$, whereas the expansion in $q$, $K_{\rm GSF}^q$, does not. (iii) Assuming the coefficients $a_n$ in the formal expansion $K=\sum_n a_n(x)\nu^n$ do not increase with $n$ (which, however, only a future calculation of higher-order GSF terms could confirm), this expansion will exhibit a fast convergence since $0 < \nu \leqslant 1/4$; the same cannot be said of the $q$-expansion.

Comparison of the GSF curves in Figs.~\ref{fig:q=1}--\ref{fig:Omega=0.022} with the NR benchmark leads us to yet another important observation. It is evident that the second-order GSF correction to $K$ (i.e. the unknown term $\propto q^2$) has an {\em opposite} sign with respect to the first-order term; namely, the second-order GSF acts to increase the rate of periastron advance. This is a new result, which illustrates the potential merit of cross-cultural comparisons of the kind advocated in this work.

{\it Summary and Discussion.} The advent of precision-NR technology allows us, for the first time, to extract accurate information about the {\em local} dynamics in binary black hole inspirals (previous studies focused primarily on asymptotic waveforms), and carry out meaningful comparisons with the results of analytic approaches to the problem. These comparisons and cross-check validations among analytic approximants and NR results are crucial for developing faithful analytic  
waveforms to be used in LIGO/Virgo searches. 

Here we focused on a particular aspect of the dynamics, namely the relativistic periastron advance. We worked in a highly relativistic regime, where the periastron advance can reach values as high as half an orbit per radial period (far greater than the meagre $\sim 43"$ per century advance of Mercury's perihelion!) We employed the invariant relation $K(\Omega_\varphi)$ as a reference for comparison, which is meaningful only in the adiabatic regime where the dissipative evolution is ``slow''. For the range of inspiral orbits covered by our NR simulations, a measure of adiabaticity is provided by $0.3\% \lesssim \dot{\Omega}_\varphi / \Omega_\varphi^2 \lesssim 1.7\%$. This suggests that inclusion of dissipative effects in the PN/EOB/GSF results would not substantially affect our conclusions. The very good agreement between the analytical and NR results at low frequency, where the error in $K_\text{NR}$ is smallest, also supports this expectation.

Our direct comparison between perturbative and full NR results is the
first of its kind. The $\mathcal{O}(q)$ GSF prediction agrees with the
NR data for small mass ratios (e.g. $q=1/8$ or $1/6$) to within a
relative difference of magnitude $\sim q^2$, as expected. This
provides an extremely strong validity test for both NR and GSF
calculations. Furthermore, the sign and magnitude of the difference
$K_\text{NR} - K^q_\text{GSF}$ give us valuable, hitherto inaccessible
information about the second-order GSF effect.

The above validation test is further reinforced by the 3PN result,
which shows a good agreement with the NR data at small
frequencies, or ``large'' separations (down to $r \sim 10 m$), especially
for comparable masses (e.g. for $q=1$ or $2/3$).
Our comparison also reaffirms the expectation that the PN approximation
performs less well in the small mass-ratio regime.

We find that the EOB(3PN) prediction of the periastron
advance is in very good agreement with the numerical one across the
entire range of mass ratios and frequencies considered. This result supports the idea that the
EOB formalism can describe the binary dynamics at {\em all} mass ratios.

Finally, we observe that the simple replacement $q\to\nu$ can extend the validity of the GSF approximation far beyond the EMRI regime. Indeed, our model $K^\nu_\text{GSF}$ agrees very well with the NR data at all frequencies, and for all mass ratios considered, including the equal mass case. This surprising result suggests that GSF calculations may very well find application in a broader range of physical problems than originally envisaged, including the modelling of intermediate mass ratio inspirals, a plausible source of gravitational waves for Advanced LIGO/Virgo \cite{Br.al.07}.

{\it Acknowledgments.} AB, ALT and AT acknowledge support from NSF Grant PHY-0903631. AB also acknowledges support from NASA grant NNX09AI81G. ALT and AT also acknowledge support from the Maryland Center for Fundamental Physics. LB acknowledges support from STFC through grant number PP/E001025/1. HP acknowledges support from the NSERC of Canada, from the Canada Research Chairs Program, and from the Canadian Institute for Advanced Research. NS acknowledges supports by the Grant-in-Aid for Scientific Research (No.~22740162) and the Global COE Program ``The Next Generation of Physics, Spun from Universality and Emergence'', from MEXT of Japan. Computations were performed by SpEC \cite{SpECwebsite} on the GPC supercomputer at the SciNet HPC Consortium. SciNet is funded by: the Canada Foundation for Innovation under the auspices of Compute Canada; the Goverment of Ontario; Ontario Research Fund -- Research Excellence; and the University of Toronto.

\bibliography{}

\end{document}